# Landau parameters for energy density functionals generated by local finite-range pseudopotentials

A. Idini[1,2], K. Bennaceur[1,3,4], and J. Dobaczewski[1,4,5,6]

[1]Department of Physics, PO Box 35 (YFL), FI-40014 University of Jyväskylä, Finland
[2]Department of Physics, University of Surrey, Guildford GU2 7XH, United Kingdom
[3]Univ Lyon, Université Lyon 1, CNRS/IN2P3, IPNL, F-69622 Villeurbanne, France
[4]Helsinki Institute of Physics, P.O. Box 64, FI-00014 University of Helsinki, Finland
[5]Department of Physics, University of York, Heslington, York YO10 5DD, United Kingdom
[6]Institute of Theoretical Physics, Faculty of Physics, University of Warsaw, ul. Hoża 69, PL-00681 Warsaw, Poland

**Abstract.** In Landau theory of Fermi liquids, the particle-hole interaction near the Fermi energy in different spin-isospin channels is probed in terms of an expansion over the Legendre polynomials. This provides a useful and efficient way to constrain properties of nuclear energy density functionals in symmetric nuclear matter and finite nuclei. In this study, we present general expressions for Landau parameters corresponding to a two-body central local regularized pseudopotential. We also show results obtained for two recently adjusted NLO and N$^2$LO parametrizations. Such pseudopotentials will be used to determine mean-field and beyond-mean-field properties of paired nuclei across the entire nuclear chart.

## 1. Introduction

The study of low-energy properties of atomic nuclei has progressed both theoretically and experimentally during the last 50 years. Thanks to systematic use of stable and exotic beams, pioneered at ISOLDE and then constantly improved, nuclear structure models can today dispose of a wealth of data for binding energies and low-energy excitations. Among other models, the nuclear density functional theory (DFT) represents the only approach that is able to describe ground-state and low-energy excited states across the entire nuclear chart. In 1958, T.H.R. Skyrme proposed an effective interaction for nuclear systems composed of contact two-body and three-body terms in the form of an expansion in relative momenta of interacting nucleons [1]. After this seminal work, other forms of nuclear effective interactions were considered with, for example, the inclusion finite-range terms [2, 3]. Furthermore, the three-body term was replaced by a density-dependent term [4], thus allowing to reproduce the empirical values of incompressibility modulus of nuclear matter and quasi-particle effective mass. However, this gave away the Hamiltonian formulation of the model and transformed the interactions into functional generators.

Over the past decade, one has seen a renewed endeavor in the development of parametrizations and analytical forms for novel nuclear energy density functionals (EDFs) and functional generators. This has been ushered by (i) the application of modern statistical tools for the fitting of density functionals to nuclear observables [5, 6, 7], (ii) the highlighting of the precision limits of present Skyrme functional forms [8], and (iii) the discovery of difficulties in beyond-mean-field calculations using most of the present functionals [9, 10, 11].

A few years ago, in the spirit of the effective theory [12], we have introduced a momentum-dependent pseudopotential that uses finite-range Gaussian regulators [13]. This approach has proven to be adequate to reproduce infinite nuclear matter properties (aside from the effective mass, due to the purely two-body nature of the considered pseudopotential), binding energies of closed-shell nuclei [14] and average pairing gaps for open-shell nuclei [15].

Adjustments of coupling constants, which define specific parametrizations of pseudopotentials, represent a challenge because of the potentially large number of parameters to be optimized, and because the finite-range derivative terms make computation of observables for finite nuclei more demanding. In this sense, calculation of infinite nuclear matter properties represents an efficient way to add constraints to the fitting procedure. In particular, Landau parameters can be constrained to their empirical values, when they are available, or can be used to avoid entering regions of instability in the parameters space (see [16] and references therein).

In the Landau-Migdal theory of Fermi liquids, properties of nuclear matter are formulated in terms of a two-body particle-hole residual interaction, calculated at the Fermi surface [17]. The obtained parametrization of the particle-hole interaction can be used for an approximate calculation of nuclear excitations [18]. Using notations which can, for example, be found in [19], we recall the following sum rules that the Landau parameters have to fulfill [20] in order to respect the Pauli exclusion principle,

$$\sum_{\ell,(\alpha)} f_\ell^{(\alpha)} = 0, \tag{1}$$

$$\sum_\ell f_\ell^{(0,1)} + f_\ell^{(1,0)} - 2f_\ell^{(1,1)} = \sum_\ell f_\ell' + g_\ell - 2g_\ell' = 0, \tag{2}$$

where indices $(\alpha) = (S,T)$ represent different spin-isospin channels. The fulfillment of these sum rules implies the avoidance of the spurious self-interaction in the particle-hole channel, usually arising from the use of density-dependent terms. A purely two-body regularized pseudopotential discussed in this article is free from self-interaction by construction, hence the sum rules have to be analytically fulfilled. A phenomenon of spurious self-pairing of similar nature appears when the normal and pairing parts of the EDF are not derived from the same effective interaction as pointed out in [9, 21].

Finally, let us recall that the correspondence between Landau parameters and bulk nuclear properties implies constrains on their values, so as to ensure the stability of homogeneous infinite nuclear matter with respect to different polarizations. This gives the Migdal stability conditions,

$$F_\ell^{(\alpha)} = N_F f_\ell^{(\alpha)} > -(2\ell + 1), \tag{3}$$

where

$$N_F = 2m^* k_F / (\pi^2 \hbar^2) \approx (m^*/m)(k_F / 204.68) \tag{4}$$

is the density of states at the Fermi surface [22].

The article is organized as follows. In section 2, we recall definitions of Landau parameters derived from a finite-range pseudopotential that may contain momentum-dependent terms. In section 3 we give explicit formulae for two pseudopotentials proposed in our recent study [15]. Results obtained for these pseudopotentials are reported in section 4 together with comparison with those obtained for the Gogny D1S effective interaction [23] and with discussion of propagated errors. Finally, conclusions are presented in section 5.

## 2. Landau parameters for a momentum-dependent central interaction

In [13], we considered a pseudopotential that is a sum of two-body terms built as products of two locality delta functions and finite-range regulator, that is, $\delta(\mathbf{r}_{13})\delta(\mathbf{r}_{24})g_a(\mathbf{r}_{12})$. Higher-order momentum-dependent terms contain differential operators of order $n$, denoted by $\hat{O}_j^{(n)}(\mathbf{k}_{12}, \mathbf{k}_{34})$, and read

$$\mathcal{V}_j^{(n)}(\mathbf{r}_1, \mathbf{r}_2; \mathbf{r}_3, \mathbf{r}_4) = \left(W_j^{(n)} \hat{1}_\sigma \hat{1}_\tau + B_j^{(n)} \hat{1}_\tau \hat{P}^\sigma - H_j^{(n)} \hat{1}_\sigma \hat{P}^\tau - M_j^{(n)} \hat{P}^\sigma \hat{P}^\tau\right)$$
$$\times \hat{O}_j^{(n)}(\mathbf{k}_{12}, \mathbf{k}_{34}) \delta(\mathbf{r}_{13}) \delta(\mathbf{r}_{24}) g_a(\mathbf{r}_{12}), \tag{5}$$

where $n$ stands for even powers of the relative momenta of incoming ($\mathbf{k}_{12} = \mathbf{k}_1 - \mathbf{k}_2$) and outgoing ($\mathbf{k}_{34} = \mathbf{k}_3 - \mathbf{k}_4$) particles at order $p = n/2$ beyond leading order (N$^p$LO).

The particle-hole residual interaction is given by the antisymmetrization of the considered pseudopotential

$$V_j^{ph(n)} = \mathcal{V}_j^{(n)}(1 - \hat{P}^x \hat{P}^\sigma \hat{P}^\tau), \tag{6}$$

and the total residual interaction is given by the sum over the various contributions at all included orders

$$V_{\text{res}}^{ph} = \sum_{n,j} \mathcal{V}_j^{(n)}(1 - \hat{P}^x \hat{P}^\sigma \hat{P}^\tau). \tag{7}$$

Making use of the Fourier transform, matrix elements in the momentum space of the particle-hole residual interaction can be written as

$$V_{\text{res}}^{ph}(\mathbf{k}_1, \mathbf{k}_2; \mathbf{k}_3, \mathbf{k}_4) = \langle \mathbf{k}_1, \mathbf{k}_2 | V_{\text{res}}^{ph} | \mathbf{k}_3 \mathbf{k}_4 \rangle = \frac{\delta(\mathbf{q} - \mathbf{q}')}{(2\pi)^3} \sum_{n,j} \mathcal{V}_j^{(n)}(q, \mathbf{k} - \mathbf{k}'), \tag{8}$$

where $\mathbf{k}_1 = \mathbf{k} + \mathbf{q}$, $\mathbf{k}_2 = \mathbf{k}'$, $\mathbf{k}_3 = \mathbf{k}$ and $\mathbf{k}_4 = \mathbf{k}' + \mathbf{q}'$, and

$$\begin{aligned}\mathcal{V}_j^{(n)}(q) &= \mathcal{V}_{(n)j}^{(0,0)}(q) + \mathcal{V}_{(n)j}^{(1,0)}(q)\, \boldsymbol{\sigma}_1 \cdot \boldsymbol{\sigma}_2 \\ &+ \mathcal{V}_{(n)j}^{(0,1)}(q)\, \boldsymbol{\tau}_1 \cdot \boldsymbol{\tau}_2 + \mathcal{V}_{(n)j}^{(1,1)}(q)\, \boldsymbol{\sigma}_1 \cdot \boldsymbol{\sigma}_2\, \boldsymbol{\tau}_1 \cdot \boldsymbol{\tau}_2,\end{aligned} \tag{9}$$

where $\boldsymbol{\sigma}_i$ and $\boldsymbol{\tau}_i$ are vectors formed with, respectively, spin and isospin Pauli matrices for particle $i$.

Landau parameters are then the coefficients of the particle-hole interaction expanded over Legendre polynomials in different spin-isospin channels $(\alpha) \equiv (S, T)$, and calculated at the Fermi surface, that is, for $|\mathbf{k}| = |\mathbf{k}'| = k_F$

$$\begin{aligned}\lim_{q \to 0} \sum_{n,j} \mathcal{V}_{(n)j}^{(\alpha)}(q, \mathbf{k} - \mathbf{k}')\bigg|_{|\mathbf{k}|=|\mathbf{k}'|=k_F} &= \sum_{n,j} D_{(n)j}^{(\alpha)}(k_F) + E_{(n)j}^{(\alpha)}(k_F, \hat{k} \cdot \hat{k}'), \\ &= \sum_\ell f_\ell^{(\alpha)}(k_F)\, P_\ell(\hat{k} \cdot \hat{k}').\end{aligned} \tag{10}$$

In the latter equation, functions $D_{(n)j}^{(\alpha)}(k_F)$ come from the direct term in Eq. (7), and depend on $k_F$ only, whereas functions $E_{(n)j}^{(\alpha)}(k_F)$ come from the exchange term and depend both on $k_F$ and on the angle between $\mathbf{k}$ and $\mathbf{k}'$.

## 3. Regularized pseudopotential

In [13, 15], a Gaussian function,

$$g_a(\mathbf{r}) = \frac{1}{(a\sqrt{\pi})^3}\, e^{-\frac{\mathbf{r}^2}{a^2}}, \tag{11}$$

was used as a regulator of the pseudopotential. In the following, we use simplified notations $\{W_0^{(0)}, B_0^{(0)}, H_0^{(0)}, M_0^{(0)}\} \equiv \{W_0, B_0, H_0, M_0\}$ for the coupling constants at LO, and $\{W_i^{(2)}, B_i^{(2)}, H_i^{(2)}, M_i^{(2)}\} \equiv \{W_i, B_i, H_i, M_i\}$ and $\{W_i^{(4)}, B_i^{(4)}, H_i^{(4)}, M_i^{(4)}\} \equiv \{W_{i+2}, B_{i+2}, H_{i+2}, M_{i+2}\}$, with $i = 1$ or 2, for the ones at NLO and N²LO, respectively.

Using Eqs. (8) and (10), Landau parameters with contributions up to second order of the pseudopotential (NLO) are given by

$$f_\ell^{(\alpha)}(k_F) = A_0^{(\alpha)} \delta_{\ell 0} + B_0^{(\alpha)}(2\ell+1) \, e^{-\frac{a^2 k_F^2}{2}} i_\ell\left(\frac{a^2 k_F^2}{2}\right)$$
$$+ \frac{k_F^2}{2}\left(A_1^{(\alpha)} + A_2^{(\alpha)}\right)(\delta_{\ell 0} - \delta_{\ell 1})$$
$$+ \frac{k_F^2}{2}\left(B_1^{(\alpha)} - B_2^{(\alpha)}\right) e^{-\frac{a^2 k_F^2}{2}}$$
$$\left[(2\ell+1)\,i_\ell\left(\frac{a^2 k_F^2}{2}\right) - \ell\, i_{\ell-1}\left(\frac{a^2 k_F^2}{2}\right) - (\ell+1)\,i_{\ell+1}\left(\frac{a^2 k_F^2}{2}\right)\right], \quad (12)$$

where $i_\ell$ is a modified spherical Bessel function of the first kind and of order $\ell$. In this expression, the coupling constants defining Landau parameters are given by

$$A_i^{(0,0)} = W_i + \tfrac{1}{2} B_i - \tfrac{1}{2} H_i - \tfrac{1}{4} M_i, \tag{13}$$
$$B_i^{(0,0)} = -\tfrac{1}{4} W_i - \tfrac{1}{2} B_i + \tfrac{1}{2} H_i + M_i, \tag{14}$$
$$A_i^{(1,0)} = \tfrac{1}{2} B_i - \tfrac{1}{4} M_i, \tag{15}$$
$$B_i^{(1,0)} = -\tfrac{1}{4} W_i + \tfrac{1}{2} H_i, \tag{16}$$
$$A_i^{(0,1)} = -\tfrac{1}{2} H_i - \tfrac{1}{4} M_i, \tag{17}$$
$$B_i^{(0,1)} = -\tfrac{1}{4} W_i - \tfrac{1}{2} B_i, \tag{18}$$
$$A_i^{(1,1)} = -\tfrac{1}{4} M_i, \tag{19}$$
$$B_i^{(1,1)} = -\tfrac{1}{4} W_i \tag{20}$$

with $i = 0$, 1 or 2.

In the case of a local pseudopotential, as discussed in [13, 15], one has $A_2^{(\alpha)} = -A_1^{(\alpha)}$ and $B_2^{(\alpha)} = -B_1^{(\alpha)}$ so that equation (12) reduces to

$$f_\ell^{(\alpha)}(k_F) = A_0^{(\alpha)} \delta_{\ell 0} + B_0^{(\alpha)}(2\ell+1) \, e^{-\frac{a^2 k_F^2}{2}} i_\ell\left(\frac{a^2 k_F^2}{2}\right)$$
$$+ k_F^2 \, B_1^{(\alpha)}(2\ell+1)\left(-\frac{1}{a}\frac{\partial}{\partial a}\right)\left[e^{-\frac{a^2 k_F^2}{2}} i_\ell\left(\frac{a^2 k_F^2}{2}\right)\right] \tag{21}$$

where the contribution at NLO is simply obtained by acting on the one at LO with the operator $-\frac{1}{a}\frac{\partial}{\partial a}$. This latter operator can be iterated $p$ times on the expression at LO to give the expression at N$^p$LO.

In the next section, we discuss results obtained for two local pseudopotentials whose parameters are given in [15]. Let us note that a contact two-body interaction had to be included in these pseudopotentials in order to avoid excessive surface pairing effects. This two-body contact interaction is a standard local contact Skyrme force

$$\mathcal{V}_\delta(\mathbf{r}_1, \mathbf{r}_2; \mathbf{r}_3, \mathbf{r}_4) = t_0\left(1 + x_0 \hat{P}^\sigma\right) \delta(\mathbf{r}_{13})\delta(\mathbf{r}_{24})\delta(\mathbf{r}_{12}), \tag{22}$$

which gives additional contributions $A_0'^{(\alpha)}$ to the Landau parameters with

$$A_0'^{(0,0)} = \tfrac{3}{4} t_0, \tag{23}$$
$$A_0'^{(1,0)} = -\tfrac{1}{4} t_0 + \tfrac{1}{2} t_0 x_0, \tag{24}$$

$$A_0'^{(0,1)} = -\tfrac{1}{4} t_0 - \tfrac{1}{2} t_0 x_0 , \tag{25}$$

$$A_0'^{(0,0)} = -\tfrac{1}{4} t_0 . \tag{26}$$

Equations (23)–(26) represent general expressions for Landau parameters obtained from interaction (22). In the case considered here, parameter $x_0$ was set to 1 so that this term was not active in the pairing channel. Taking into account this contact term and considering the fact that the finite-range part of the pseudopotential is local, expressions for Landau parameters discussed in the next section are given by

$$f_\ell^{(\alpha)}(k_F) = (A_0'^{(\alpha)} + A_0^{(\alpha)})\delta_{\ell 0} + B_0^{(\alpha)}(2\ell + 1)\,\mathrm{e}^{-\frac{a^2 k_F^2}{2}} i_\ell\left(\frac{a^2 k_F^2}{2}\right)$$
$$+ (2\ell + 1)\, k_F^2 \sum_{i=1}^{2} B_{2i-1}^{(\alpha)} \left(-\frac{1}{a}\frac{\partial}{\partial a}\right)^i \left[\mathrm{e}^{-\frac{a^2 k_F^2}{2}} i_\ell\left(\frac{a^2 k_F^2}{2}\right)\right] . \tag{27}$$

## 4. Results in symmetric nuclear matter

In the following we consider the NLO and N$^2$LO local regularized pseudopotentials as in Eq. (5), supplemented by the contact term given by Eq. (22), and we show Landau parameters evaluated for two recent parametrizations REG2c.161026 and REG4c.161026 [15], respectively. Using Eq. (27), we calculate the $\ell = 0, 1, 2$ Landau parameters as functions of the nuclear-matter density as depicted in Figs. 1 and 2. For comparison, in Fig. 3 we show results obtained for the Gogny D1S force [23]. Landau parameters at saturation densities, with corresponding theoretical errors, are summarized in Table 1.

We begin by recalling relations between Landau parameters and infinite-nuclear-matter properties. For the effective mass one has,

$$m^* = m\left(1 + \frac{F_1}{3}\right), \tag{28}$$

and for REG2c.161026 and REG4c.161026 we obtain values of $m^* = (0.4076 \pm 0.0009)\,m$ and $(0.4061 \pm 0.0011)\,m$, respectively. These can be used to consistently calculate densities of states $N_F$, Eq. (4), which gives $0.02655 \pm 0.00006$ and $0.02655 \pm 0.00007\,\mathrm{MeV}^{-1}\,\mathrm{fm}^{-3}$. As discussed in [15], small values of effective masses are obtained for pure two-body pseudopotentials. This implies that some Landau parameters like $F_0$ and $F_1$ are not consistent with literature references, e.g., [24] provides limits of $-0.51 < F_0 < 0.21$ and $-1.07 < F_1 < -0.89$.

The incompressibility modulus and symmetry energy coefficient of nuclear matter can also be related to Landau parameters as [25],

$$K_\infty = \frac{3\hbar^2 k_F^2}{m^*}(1 + F_0), \tag{29}$$

$$J = \frac{\hbar^2 k_F^2}{6m^*}(1 + F_0'), \tag{30}$$

giving $K_\infty = 229.8 \pm 4.8$ MeV and $J = 31.96 \pm 4.71$ MeV for REG2c.161026 and $K_\infty = 230.0 \pm 5.6$ MeV and $J = 31.95 \pm 5.41$ MeV for REG4c.161026, consistently with the constraints imposed on these properties during the fitting procedure [15].

An analysis of the spin and spin-isospin response of nuclei performed in [26, 27] leads to the recommended values of $G_0 \simeq 0.15$ and $G'_0 \simeq 0.3$. In fact, the spin-isospin channel of the particle-hole interaction is related to the Gamow-Teller distribution, and for contact interactions can be easily parametrized through Landau parameters as [28],

$$C_1^s = \frac{\pi^2 \hbar^2}{2m^* k_F}(G'_0 + G'_1), \tag{31}$$

where $C_1^s$ is the coupling constant defining the isovector spin-spin interaction [29]. Although our NLO and N$^2$LO pseudopotentials are finite range, values extracted from Landau parameters, Eq. (31), that is, $C_1^s = 135.8 \pm 0.4$ and $148.4 \pm 0.8$ MeV fm$^3$, respectively, compare quite favorably with that of $\sim 120$ MeV fm$^3$ quoted in [28].

Since our regularized pseudopotentials supplemented with the contact two-body terms are pure two-body EDF generators that do not introduce any spuriousities, the Migdal sum rules of Eqs. (1) and (2) are exactly obeyed.

All propagated errors quoted in this study result from the statistical error analysis performed for the parameters of the REG2c.161026 and REG4c.161026 pseudopotentials. We made use of the covariance matrices determined in [15] along the lines delineated in [6]. That is, first the Hessian matrix built of derivatives of the scaled penalty function with respect to parameters of the pseudopotential was calculated. Second, the covariance matrix was obtained as the inverse of the Hessian matrix, while keeping only a given number of its largest eigenvalues.

In Fig. 4, we see that the propagated errors of the Landau parameters are qualitatively similar when going from NLO to N$^2$LO. Errors of $F_0$, $F'_0$, and $F_1$ form a plateau after keeping 4 or 5 eigenvalues. This means that, as we expect, these parameters are already well constrained by the current penalty function. They are, in fact, determined by constraining $K_\infty$ and $J$, whereas for purely two–body functional generators, the effective mass is always $\sim 0.4m$ [15]. On the contrary errors of $F'_1$, $G_1$, $G'_1$ seem to increase up to the inclusion of the very last eigenvalue. This means that the current penalty function does not carry information concerning the properties of these particle-hole channels, and that, by including these Landau parameters into the fitting procedure, future parametrizations might be better constrained.

## 5. Conclusions

We presented general expressions for Landau parameters expressed in terms of Fourier transform of a general two-body central regularized pseudopotential, and we specified them to the case of local NLO and N$^2$LO pseudopotentials. We also showed numerical values of Landau parameters derived for two recent parametrizations REG2c.161026 and REG4c.161026.

In the future, the results obtained here will be used to improve empirical properties of nuclear energy density functionals. Specifically, we showed that constraining Landau parameters that correspond to spin-dependent channels is a promising route towards better determination of the EDF coupling constants. Work is now going on towards developing spuriousity-free functionals that are consistent in particle-hole and in particle-particle channels and that lead to more realistic values of the effective mass.

|        | NLO          | N$^2$LO      | D1S    | [25]  | [26, 27] | [28] | [30]     |
|--------|--------------|--------------|--------|-------|----------|------|----------|
| $F_0$  | $-0.576(12)$ | $-0.578(14)$ | $-0.369$ | 0.1   |          |      | 0±0.16   |
| $F'_0$ | $0.061(9)$   | $0.056(10)$  | $0.743$  | 0.7   |          |      | 0.64     |
| $G_0$  | $0.151(10)$  | $0.189(7)$   | $0.466$  | 1.15  | 0.15     |      | 0.16     |
| $G'_0$ | $0.982(2)$   | $1.044(5)$   | $0.631$  | 1.45  | 0.3      | 1.2  | 0.72     |
| $F_1$  | $-1.777(4)$  | $-1.782(5)$  | $-0.909$ | -0.6  |          |      |          |
| $F'_1$ | $0.766(5)$   | $0.553(3)$   | $0.470$  | 0.5   |          |      |          |
| $G_1$  | $0.015(2)$   | $-0.018(3)$  | $-0.185$ | 0.3   |          |      |          |
| $G'_1$ | $-0.01452(15)$ | $0.10868(98)$ | $0.610$ | 0.3  |          | 0.2  |          |

**Table 1.** Landau parameters calculated at saturation densities of $\rho_{\text{sat}} = 0.1599\,\text{fm}^{-3}$ (regularized pseudopotential at NLO, REG2c.161026), $\rho_{\text{sat}} = 0.1601\,\text{fm}^{-3}$ (regularized pseudopotential at N$^2$LO, REG4c.161026) or $0.1633\,\text{fm}^{-3}$ (Gogny D1S force), compared with empirical values quoted in [25]–[30].

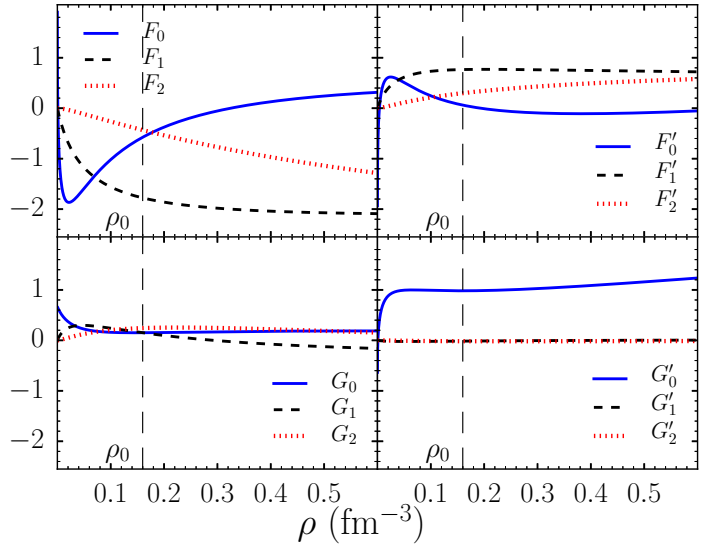

**Figure 1.** (Color online) Landau parameters for the REG2c.161026 pseudopotential as functions of density, determined for different spin-isospin channels and for $\ell = 0$ (solid line), $\ell = 1$ (dashed line), and $\ell = 2$ (dotted line). Saturation density $\rho_0 = 0.16\,\text{fm}^{-3}$ is denoted by a dashed vertical line.

## Acknowledgments

This work was supported by the Academy of Finland and University of Jyväskylä within the FIDIPRO program, by the Royal Society and Newton Fund under the Newton

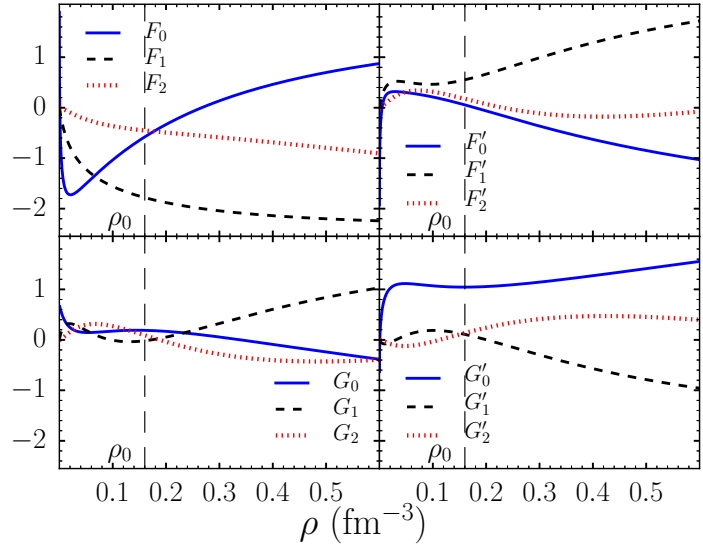

**Figure 2.** (Color online) Same as in Fig. 1 but for the REG4c.161026 pseudopotential.

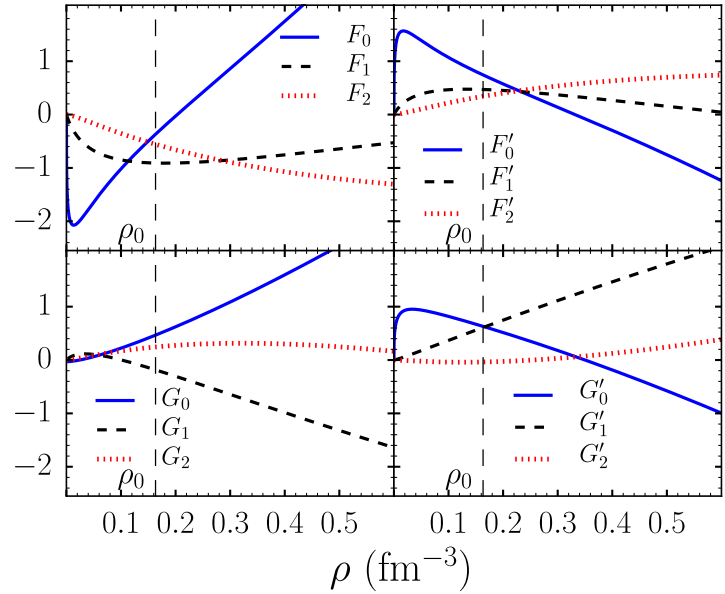

**Figure 3.** (Color online) Same as in Fig. 1 but for the Gogny force D1S [23].

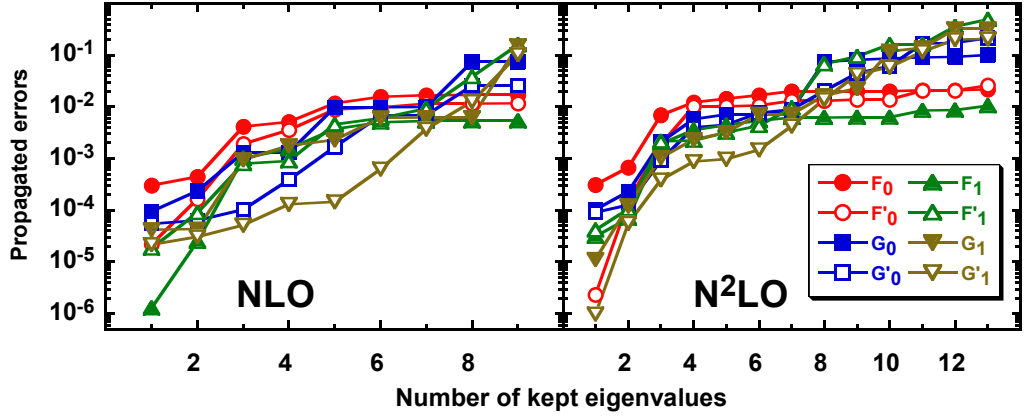

**Figure 4.** (Color online) Propagated errors of Landau Parameters $F_l^{(\alpha)}$ for all isospin channels and $l = 0$ and 1 in absolute scale as a function of smallest eigenvalues kept in the spectrum of the covariance matrix. Left panel: calculated with REG2c.161026 NLO local pseudopotential. Right panel: calculated with REG4c.161026 N$^2$LO local pseudopotential.

**References**

[1] Skyrme T H R 1959 *Nuclear Physics* **9** 615 – 634 ISSN 0029-5582 URL http://www.sciencedirect.com/science/article/pii/0029558258903456
[2] Brink D and Boeker E 1967 *Nuclear Physics A* **91** 1 – 26 ISSN 0375-9474 URL http://www.sciencedirect.com/science/article/pii/0375947467904460
[3] Dechargé J and Gogny D 1980 *Phys. Rev. C* **21**(4) 1568–1593
[4] Vautherin D and Brink D M 1972 *Phys. Rev. C* **5**(3) 626–647 URL http://link.aps.org/doi/10.1103/PhysRevC.5.626
[5] Toivanen J, Dobaczewski J, Kortelainen M and Mizuyama K 2008 *Phys. Rev. C* **78**(3) 034306 URL http://link.aps.org/doi/10.1103/PhysRevC.78.034306
[6] Dobaczewski J, Nazarewicz W and Reinhard P G 2014 *Journal of Physics G: Nuclear and Particle Physics* **41** 074001 URL http://stacks.iop.org/0954-3899/41/i=7/a=074001
[7] Ireland D G and Nazarewicz W 2015 *Journal of Physics G: Nuclear and Particle Physics* **42** 030301 URL http://stacks.iop.org/0954-3899/42/i=3/a=030301
[8] Kortelainen M, McDonnell J, Nazarewicz W, Olsen E, Reinhard P G, Sarich J, Schunck N, Wild S M, Davesne D, Erler J and Pastore A 2014 *Phys. Rev. C* **89**(5) 054314 URL http://link.aps.org/doi/10.1103/PhysRevC.89.054314
[9] Anguiano M, Egido J and Robledo L 2001 *Nuclear Physics A* **696** 467 – 493 ISSN 0375-9474 URL http://www.sciencedirect.com/science/article/pii/S0375947401012192
[10] Dobaczewski J, Stoitsov M V, Nazarewicz W and Reinhard P G 2007 *Phys. Rev. C* **76**(5) 054315 URL http://link.aps.org/doi/10.1103/PhysRevC.76.054315
[11] Lacroix D, Duguet T and Bender M 2009 *Phys. Rev. C* **79**(4) 044318 URL http://link.aps.org/doi/10.1103/PhysRevC.79.044318
[12] Dobaczewski J, Bennaceur K and Raimondi F 2012 *Journal of Physics G: Nuclear and Particle Physics* **39** 125103 URL http://stacks.iop.org/0954-3899/39/i=12/a=125103
[13] Raimondi F, Bennaceur K and Dobaczewski J 2014 *Journal of Physics G: Nuclear and Particle Physics* **41** 055112 URL http://stacks.iop.org/0954-3899/41/i=5/a=055112
[14] Bennaceur K, Dobaczewski J and Raimondi F 2014 *EPJ Web of Conferences* **66** 02031 URL http://dx.doi.org/10.1051/epjconf/20146602031
[15] Bennaceur K, Idini A, Dobaczewski J, Dobaczewski P, Kortelainen M and Raimondi F submitted to *Journal of Physics G: Nuclear and Particle Physics* (*Preprint* arXiv:1611.09311)
[16] Davesne D, Pastore A and Navarro J 2013 *Journal of Physics G: Nuclear and Particle Physics* **40** 095104 URL http://stacks.iop.org/0954-3899/40/i=9/a=095104
[17] Migdal A 1967 *Theory of Finite Fermi Systems And Applications to Atomic Nuclei* Interscience monographs and texts in physics and astronomy, v. 19 (Interscience Publishers)
[18] Ring P and Schuck P 2004 *The Nuclear Many-Body Problem* Physics and astronomy online library (Springer) ISBN 9783540212065
[19] Bender M, Dobaczewski J, Engel J and Nazarewicz W 2002 *Phys. Rev. C* **65**(5) 054322 URL http://link.aps.org/doi/10.1103/PhysRevC.65.054322
[20] Bäckman S O, Brown G, Klemt V and Speth J 1980 *Nuclear Physics A* **345** 202 – 220 ISSN 0375-9474 URL http://www.sciencedirect.com/science/article/pii/0375947480904194
[21] Bender M, Duguet T and Lacroix D 2009 *Phys. Rev. C* **79**(4) 044319 URL http://link.aps.org/doi/10.1103/PhysRevC.79.044319
[22] Pastore A, Davesne D and Navarro J 2015 *Physics Reports* **563** 1 – 67 ISSN 0370-1573 linear response of homogeneous nuclear matter with energy density functionals URL http://www.sciencedirect.com/science/article/pii/S0370157314003998
[23] Berger J, Girod M and Gogny D 1991 *Computer Physics Communications* **63** 365 – 374 ISSN 0010-4655 URL http://www.sciencedirect.com/science/article/pii/001046559190263K
[24] Caillon J C, Gabinski P and Labarsouque J 2002 *Journal of Physics G: Nuclear and Particle Physics* **28** 189 URL http://stacks.iop.org/0954-3899/28/i=1/a=314

[25] Bäckman S O, Jackson A and Speth J 1975 *Physics Letters B* **56** 209 – 211 ISSN 0370-2693 URL http://www.sciencedirect.com/science/article/pii/0370269375903767
[26] Wakasa T, Ichimura M and Sakai H 2005 *Phys. Rev. C* **72**(6) 067303 URL http://link.aps.org/doi/10.1103/PhysRevC.72.067303
[27] Roca-Maza X, Colò G and Sagawa H 2012 *Phys. Rev. C* **86**(3) 031306 URL http://link.aps.org/doi/10.1103/PhysRevC.86.031306
[28] Bender M, Dobaczewski J, Engel J and Nazarewicz W 2002 *Phys. Rev. C* **65**(5) 054322 URL http://link.aps.org/doi/10.1103/PhysRevC.65.054322
[29] Perlińska E, Rohoziński S G, Dobaczewski J and Nazarewicz W 2004 *Phys. Rev. C* **69**(1) 014316 URL http://link.aps.org/doi/10.1103/PhysRevC.69.014316
[30] Speth J, Krewald S and Grümmer F 2002 *Theory of Finite Fermi Systems – The Stony Brook Jülich Interaction* From Nuclei to Stars Festschrift in Honor of Gerald E Brown (World Scientific) URL http://www.worldscientific.com/worldscibooks/10.1142/8017